\documentstyle[12pt]{article}

\def\hybrid{\topmargin -20pt    \oddsidemargin 0pt
        \headheight 0pt \headsep 0pt
        \textwidth 6.25in       
        \textheight 9.5in       
        \marginparwidth .875in
        \parskip 5pt plus 1pt   \jot = 1.5ex}

\hybrid

\def\baselinestretch{1.2}

\catcode`\@=11

\def\marginnote#1{}
\newcount\hour
\newcount\minute
\newtoks\amorpm
\hour=\time\divide\hour by60
\minute=\time{\multiply\hour by60 \global\advance\minute by-\hour}
\edef\standardtime{{\ifnum\hour<12 \global\amorpm={am}%
        \else\global\amorpm={pm}\advance\hour by-12 \fi
        \ifnum\hour=0 \hour=12 \fi
        \number\hour:\ifnum\minute<10 0\fi\number\minute\the\amorpm}}
\edef\militarytime{\number\hour:\ifnum\minute<10 0\fi\number\minute}

\def\draftlabel#1{{\@bsphack\if@filesw {\let\thepage\relax
   \xdef\@gtempa{\write\@auxout{\string
      \newlabel{#1}{{\@currentlabel}{\thepage}}}}}\@gtempa
   \if@nobreak \ifvmode\nobreak\fi\fi\fi\@esphack}
        \gdef\@eqnlabel{#1}}
\def\@eqnlabel{}
\def\@vacuum{}
\def\draftmarginnote#1{\marginpar{\raggedright\scriptsize\tt#1}}

\def\draft{\oddsidemargin -.5truein
        \def\@oddfoot{\sl preliminary draft \hfil
        \rm\thepage\hfil\sl\today\quad\militarytime}
        \let\@evenfoot\@oddfoot \overfullrule 3pt
        \let\label=\draftlabel
        \let\marginnote=\draftmarginnote
   \def\@eqnnum{(\theequation)\rlap{\kern\marginparsep\tt\@eqnlabel}%
\global\let\@eqnlabel\@vacuum}  }

\def\preprint{\twocolumn\sloppy\flushbottom\parindent 2em
        \leftmargini 2em\leftmarginv .5em\leftmarginvi .5em
        \oddsidemargin -.5in    \evensidemargin -.5in
        \columnsep .4in \footheight 0pt
        \textwidth 10.in        \topmargin  -.4in
        \headheight 12pt \topskip .4in
        \textheight 6.9in \footskip 0pt
        \def\@oddhead{\thepage\hfil\addtocounter{page}{1}\thepage}
        \let\@evenhead\@oddhead \def\@oddfoot{} \def\@evenfoot{} }

\def\numberbysection{\@addtoreset{equation}{section}
        \def\theequation{\thesection.\arabic{equation}}}

\def\underline#1{\relax\ifmmode\@@underline#1\else
        $\@@underline{\hbox{#1}}$\relax\fi}

\def\titlepage{\@restonecolfalse\if@twocolumn\@restonecoltrue\onecolumn
     \else \newpage \fi \thispagestyle{empty}\c@page\z@
        \def\thefootnote{\fnsymbol{footnote}} }

\def\endtitlepage{\if@restonecol\twocolumn \else \newpage \fi
        \def\thefootnote{\arabic{footnote}}
        \setcounter{footnote}{0}}  

\catcode`@=12
\relax

\def\figcap{\section*{Figure Captions\markboth
        {FIGURECAPTIONS}{FIGURECAPTIONS}}\list
        {Figure \arabic{enumi}:\hfill}{\settowidth\labelwidth{Figure
999:}
        \leftmargin\labelwidth
        \advance\leftmargin\labelsep\usecounter{enumi}}}
 \relax
\def\tablecap{\section*{Table Captions\markboth
        {TABLECAPTIONS}{TABLECAPTIONS}}\list
        {Table \arabic{enumi}:\hfill}{\settowidth\labelwidth{Table
999:}
        \leftmargin\labelwidth
        \advance\leftmargin\labelsep\usecounter{enumi}}}
 \relax
\def\reflist{\section*{References\markboth
        {REFLIST}{REFLIST}}\list
        {[\arabic{enumi}]\hfill}{\settowidth\labelwidth{[999]}
        \leftmargin\labelwidth
        \advance\leftmargin\labelsep\usecounter{enumi}}}
 \relax
\makeatletter
\newcounter{pubctr}
\def\publist{\@ifnextchar[{\@publist}{\@@publist}}
\def\@publist[#1]{\list
        {[\arabic{pubctr}]\hfill}{\settowidth\labelwidth{[999]}
        \leftmargin\labelwidth
        \advance\leftmargin\labelsep
        \@nmbrlisttrue\def\@listctr{pubctr}
        \setcounter{pubctr}{#1}\addtocounter{pubctr}{-1}}}
\def\@@publist{\list
        {[\arabic{pubctr}]\hfill}{\settowidth\labelwidth{[999]}
        \leftmargin\labelwidth
        \advance\leftmargin\labelsep
        \@nmbrlisttrue\def\@listctr{pubctr}}}
 \relax
\makeatother
\newskip\humongous \humongous=0pt plus 1000pt minus 1000pt

\newif\ifdtup

\relax

\def\be{\begin{equation}}
\def\ee{\end{equation}}
\def\ba{\begin{eqnarray}}
\def\ea{\end{eqnarray}}

\def\del{\partial}

\def\r{\rho}
\def\a{\alpha}

\def\b{\beta}

\def\g{\gamma}
\def\G{\Gamma}
\def\d{\delta}
\def\D{\Delta}
\def\e{\epsilon}

\def\th{\theta}

\def\m{\mu}
\def\n{\nu}

\def\l{\lambda}

\def\s{\sigma} 
\def\S{\Sigma}

\def\no{\noindent}

\def\qq{\qquad}

\def\IR{\relax{\rm I\kern-.18em R}}

\def \ha {{1\over 2}}

\def \ov {\over}

\def\diag{{\rm diag}}

\def\IR{\relax{\rm I\kern-.18em R}}
\def\inv{^{\raise.15ex\hbox{${\scriptscriptstyle -}$}\kern-.05em 1}}

\def\PL{Poisson--Lie T-duality}

\def\tL{{\tilde L}}

\begin{document}

\renewcommand{\theequation}{\thesection.\arabic{equation}}

\newcommand{\beq}{\begin{equation}}
\newcommand{\eeq}[1]{\label{#1}\end{equation}}
\newcommand{\ber}{\begin{eqnarray}}
\newcommand{\eer}[1]{\label{#1}\end{eqnarray}}
\newcommand{\eqn}[1]{(\ref{#1})}
\begin{titlepage}
\begin{center}

\hfill CERN-TH/99-112\\
\hfill hep--th/9904188\\

\vskip .8in

{\large \bf Duality-invariant class of two-dimensional field theories}

\vskip 0.6in

{\bf Konstadinos Sfetsos}
\vskip 0.1in
{\em Theory Division, CERN\\
     CH-1211 Geneva 23, Switzerland\\
{\tt sfetsos@mail.cern.ch}}\\
\vskip .2in

\end{center}

\vskip .6in

\centerline{\bf Abstract }

\vskip 0,2cm
\no
We construct a new class of two-dimensional field theories with target spaces 
that are finite multiparameter deformations of the usual coset 
$G/H$-spaces. They arise naturally, 
when certain models, related by \PL, develop a local gauge invariance 
at specific points of their classical moduli space.
We show that canonical equivalences in this context can be 
formulated in loop space in terms of parafermionic-type algebras with a 
central extension.
We find that the corresponding generating functionals 
are non-polynomial in the derivatives of the fields with respect to
the space-like variable.
After constructing models with three- and two-dimensional targets, 
we study renormalization group flows in this context.
In the ultraviolet, in some cases, 
the target space of the theory reduces to a coset space or 
there is a fixed point where the theory becomes free.

\vskip 4cm
\noindent
CERN-TH/99-112\\
April 1999\\
\end{titlepage}
\vfill
\eject

\def\baselinestretch{1.2}
\baselineskip 16 pt
\noindent

\def\tT{{\tilde T}}
\def\tg{{\tilde g}}
\def\tL{{\tilde L}}


\section{Introduction}

Cosets $G/H$ as target spaces in 2-dim field theories have been 
extensively studied in the literature, as they 
provide examples of spaces other than group manifolds, which give rise to 
integrable models.\footnote{Examples include the $O(N)$ \cite{ON1}, the 
principal chiral \cite{PC} and the Gross--Neveu models \cite{GN1}, 
for which the 
complete $S$-matrix was found through the existence of higher-spin-conserved 
currents that lead to its factorization property.
Building on work in \cite{PWW}, comparison between the $S$-matrix 
results and those obtained by 
perturbative techniques in the ultraviolet (UV) regime was made 
for the $O(N)$ $\s$-model \cite{ON2}, the Principal Chiral models for
$SU(N)$ \cite{SUN}, $SO(N)$ and 
$Sp(N)$ \cite{sosp} and the $O(N)$ Gross--Neveu model \cite{GN2}, 
finding perfect agreement.}
It is always of interest to find integrable 
deformations \cite{rajeev}--\cite{sausage}
of such models and if possible classify them. In the ordinary (undeformed)
coset models one starts with the usual Wess--Zumino action for a group, with 
Lagrangian density proportional to ${\rm Tr}(\del_i g\inv \del^i g)$, and then 
restricts the trace to the coset space only. Hence, this construction, but
not the corresponding models, is quite trivial. 
Having in mind 2-dim field theories, with targets 
spaces representing continuous deformations of the latter coset 
spaces, we need
models with non-trivial moduli as a starting point. Such an example was
considered in \cite{PLsfe3}; we present in this paper
the generalization of this to a class of theories.

We found natural to start, in section 2,
with 2-dim models related by \PL\ \cite{KliSevI}, 
since these have indeed a non-trivial moduli space
and, moreover, their classical equivalence has been established
\cite{PLsfe1, PLsfe2}.
Also, in some examples, there are hints that point towards the classical 
equivalence promoted into a quantum one
at 1-loop in perturbation theory \cite{PLsfe3,toappp}.
We will show that in some points in this moduli space a local (gauge-like)
invariance is developing. Hence, at these points the configuration
space is lower-dimensional and we discover in a unifying manner 
spaces that are deformations of the usual coset spaces.
In addition, as a byproduct, we will obtain duals of these models that 
are classically canonically equivalent to them as 2-dim field theories.
This equivalence is encoded in infinite-dimensional current algebras of 
the parafermionic type that we construct. We derive these from the 
infinite-dimensional algebras with a central extension, 
which were found in the
proof of canonical equivalence of the \PL-related models in \cite{PLsfe2}.
The corresponding generating functionals have the new feature that they are
not linear in the derivatives of the fields with respect to the 
space-like variable. This is in contrast with the cases of 
Abelian duality \cite{AALcan},
non-Abelian duality in Principal Criral \cite{zacloz,loz} and 
more general \cite{sfepara} models, 
as well as for \PL\ (and its possible generalizations) \cite{PLsfe2,PLsfe3}. 
They are, instead, non-polynomial functions of these derivatives. 
Many of these aspects are explicitly demonstrated in section 3, with a 
particular example. 
In section 4 we discuss the renormalization group (RG) 
flow in this context.
As in \cite{PLsfe3}, we emphasize that 
taking the classical limit that leads to the lower-dimensional 
models and then studying the RG flow does not necessarily imply that this
limit would correspond to a fixed point of the RG flow, i.e. the 
two procedures do not commute. There is, however, a particular domain 
in parameter space, where for a wide range of energies in the UV, 
the description is effectively perturbative 
with a UV-stable fixed point corresponding to the
point where the gauge invariance develops. Then the 
model becomes effectively a two-dimensional one.

We end the paper with section 5, 
containing concluding remarks and a discussion 
on future directions of this research. We have also written an appendix, where
some mathematical aspects of our proofs are worked out explicitly.

\section{General formulation}
\setcounter{equation}{0}

In this section we first show how new 2-dim field theories, with 
target spaces representing deformed coset spaces, arise in the
context of \PL-related $\s$-models. We then present a duality-invariant 
formulation and show that canonical equivalences are encoded into algebras
of the parafermionic-type in loop space.

\subsection{Formulation using \PL-related $\s$-models}

The form of 2-dim $\s$-model actions related by \PL\ 
(in the absence of spectator fields) is \cite{KliSevI}
\be
S= {1\ov 2 \l} \int E_{AB} L^A_M L^B_N \del_+ X^M \del_- X^N~ , 
~~~~ E= (E_0\inv + \Pi)\inv~ ,
\label{action1}
\ee
and 
\be
\tilde S= {1\ov 2 \l} \int \tilde E^{AB} \tL_{AM} \tL_{BN} \del_+
 \tilde X^M \del_- \tilde X^N~ , 
~~~~ \tilde E=(E_0 + \tilde \Pi)\inv~ .
\label{action2}
\ee
The field variables in \eqn{action1} are
$X^M$, $\m=1,2,\dots  ,d_G$ and parametrize an element
$g$ of a group $G$.
We also introduce representation matrices
$\{T_A\}$, with $A=1,2,\dots, d_G$ and
the components of the left-invariant Maurer--Cartan forms $L^A_M$.
The light-cone coordinates on the 2-dim space-time are
$x^\pm =\ha (t \pm x)$, whereas $\l$ denotes the overall coupling constant,
which is assumed to be positive. 
Similarly, for \eqn{action2} the 
field variables are $\tilde X^M$, where
$\tilde X^\m$, $\m=1,2,\dots  ,d_G$, parametrize a different group 
$\tilde G$, whose dimension is, however, equal to that of $G$.
Accordingly, we introduce a different set 
of representation matrices $\{\tilde T^A\}$, with $A=1,2,\dots, d_G$, and
the corresponding components of the left-invariant Maurer--Cartan forms 
$\tilde  L_{AM}$. In \eqn{action1} and \eqn{action2},
$E_0$ is a constant $d_G \times d_G$ matrix, whereas 
$\Pi$ and $\tilde \Pi$ are antisymmetric matrices with the 
same dimension as $E_0$, but they depend on the variables $X^M$ and $\tilde
X^M$ via the corresponding group elements $g$ and $\tilde g$.
They are defined as \cite{KliSevI}
\be
\Pi^{AB} = b^{CA} a_C{}^B ~ , ~~~~~~ 
\tilde \Pi_{AB} = \tilde b_{CA} \tilde a^C{}_B ~ ,
\label{pipi}
\ee
where the matrices $a(g)$, $b(g)$ are constructed using
\be
g\inv T_A g = a_A{}^B T_B~ ,~~~~ g\inv \tilde T^A g = 
b^{AB} T_B +  (a\inv)_B{}^A \tilde T^B~ ,
\label{abpi}
\ee
and similarly for $\tilde a(\tilde g)$ and $\tilde b(\tilde g)$.
Consistency restricts these to obey
\be
a(g\inv) = a\inv(g)~ ,~~~~ b^T(g)= b(g\inv)~ ,~~~~
\Pi^T(g) = - \Pi(g) ~ ,
\label{conss}
\ee 
and similarly for the tilded ones. 
There is also a bilinear invariant 
$\langle{\cdot|\cdot \rangle}$ with the various generators obeying 
\be
\langle{T_A|T_B\rangle}= \langle{\tilde T^A|\tilde T^B \rangle}= 0~ ,
~~~~ \langle{T_A|\tilde T^B \rangle} = \d_A{}^B ~ .
\label{bili}
\ee
Finally, we note that 
the choice of possible groups $G$ and $\tilde G$ is restricted by the 
fact that \cite{KliSevI} their corresponding Lie algebras must form a pair 
of maximally isotropic subalgebras into which the Lie algebra of a larger 
group $D$, known as the Drinfeld double, can be decomposed \cite{alemal}.

Let us consider two subgroups $H\in G$ and $\tilde H\in \tilde G$ with
$d_H = d_{\tilde H}$.
Accordingly we split the Lie-algebra indices as $A=(a,\a)$,
where Latin and Greek indices refer to subgroup and coset spaces, respectively.
Then we may separate the various matrices appearing in (\ref{action1})
and \eqn{action2}
into blocks as
\be
(E_0\inv)^{AB}= \pmatrix{E_0^{ab} & E_2^{a\b}\cr
E_3^{\a b} & E_1^{\a\b}\cr }\ ,
\label{bllo}
\ee
and 
\be
\Pi^{AB}= \pmatrix{\Pi_0^{ab} & \Pi_2^{a\b}\cr
-\Pi_2^{b \a } & \Pi_1^{\a\b}}\ ,\qq
\tilde \Pi_{AB}= \pmatrix{(\tilde \Pi_0)_{ab} & (\tilde \Pi_2)_{a\b}\cr
-(\tilde \Pi_2)_{b \a } & (\tilde \Pi_1)_{\a\b}}\ .  
\label{bllo1}
\ee

\no
We would like to take a limit in the model (\ref{action1}) and its dual 
(\ref{action2}) such that the number of fields $X^M$ (and ${\tilde X}^M$)
is reduced by $d_H$. We would call the remaining 
variables by $X^\m$ (and $\tilde X^\m$) with $\m=1,2,\dots , d_{G/H}$.
Consider the limit
\be
E_0^{ab}\to \infty\quad \Longleftrightarrow \quad (E_0\inv)_{ab}\to 0\ ,
\label{eoin}
\ee
in a uniform way for all matrix elements. This means that ratios of matrix 
elements remain constant in this limit. Using (\ref{bllo}) we find that
in the limit \eqn{eoin} 
\be
E_0 \approx \pmatrix{0& 0 \cr 0 & E_1\inv} \ .
\label{eoap}
\ee 
Then, the actions (\ref{action1}) and (\ref{action2}) take the form 
\be
S= {1\ov 2 \l} \int \S_{\a\b} L^\a_\m L^\b_\n \del_+ X^\m \del_- X^\n~ , 
\qq \S = (E_1  + \Pi_1)\inv~ ,
\label{actn1}
\ee
and 
\be
\tilde S= {1\ov 2 \l} \int \tilde \S^{AB} \tL_{A\m} \tL_{B\n} \del_+
 \tilde X^\m \del_- \tilde X^\n~ , 
\qq \tilde \S=\pmatrix{\tilde \Pi_0 & \tilde \Pi_2 \cr -\tilde \Pi_2
& E_1\inv +\tilde \Pi_1 }\inv~ .
\label{actn2}
\ee
Notice that in (\ref{actn1}) $\S_{\a\b}$ are elements of a $d_{G/H}\times
d_{G/H}$ matrix, whereas in (\ref{actn2}) $\tilde \S^{AB}$ are elements
of a $d_G\times d_G$ one. 
We have anticipated that the number of variables in (\ref{actn1}) and 
(\ref{actn2}) has been reduced to $d_{G/H}$ upon taking the limit 
(\ref{eoin}). However, this does not happen automatically, but depends 
on whether or not certain conditions, as we will next prove, are fulfilled. 
In order to reduce the dimensionality of (\ref{action1}) we should prove that,
after taking the limit (\ref{eoin}), a local gauge invariance develops,
which suffices to gauge-fix $d_H$ 
degrees of freedom in the actions \eqn{action1} and similarly for 
\eqn{action2}.
For \eqn{action1} consider the transformation
\be
g\to g h \ ,\qq h(x^+,x^-) \in H\ .
\label{trfin}
\ee
In its infinitesimal form it reads
$\d g = i g \e_a T^a$. We may show that this induces the following 
transformations: 
\be
\d L_\pm^\a = f_{b\g}{}^\a \e^b L_\pm^\g\ ,\qq \d X^M = L^M_a \e^a\ .
\label{indd}
\ee
Using these and the relation (A.6) of \cite{PLsfe2}, 
specialized for coset space 
indices\footnote{Our symmetrization and antisymmetrization conventions
are $(ab)=ab+ba$ and $[ab]=ab-ba$.}
\be
L^M_a \del_M \Pi_1^{\g\d} = - \tilde f^{\g\d}{}_a 
- f_{a \b}{}^{[\g} \Pi_1^{\d]\b}\ ,
\label{hgj}
\ee
we may prove that (\ref{actn1}) is invariant under the gauge transformation
(\ref{trfin}), provided that the following condition holds:
\be
\tilde f^{\a\b}{}_c + f_{c\g}{}^\a E_1^{\g\b} + f_{c\g}{}^\b E_1^{\a\g}= 0\ ,
\label{rel1}
\ee
or equivalently 
\be
f_{c\g}{}^{(\a} S^{\b)\g} = 0\ , \qq \tilde f^{\a\b}{}_c + f_{c\g}{}^{[\b} 
A^{\a]\g} = 0\ ,
\label{rel2}
\ee
where we have denoted
by $S^{\a\b}$ and $A^{\a\b}$ the symmetric and antisymmetric 
parts of the matrix $E_1^{\a\b}$. 
When the conditions (\ref{rel1}) are satisfied then we may gauge-fix 
$d_H$ parameters in the group element $g\in G$.
The most efficient way is to parametrize the group element $g\in G$ as 
$g=\kappa h$, where $h\in H$ and $\kappa\in G/H$, and then set $h=I$. 
It can be easily seen that this completely fixes the gauge freedom. 

There are $d_{G/H}^2 d_H$ algebraic 
conditions in \eqn{rel1} for the $d_{G/H}^2$ elements
of the matrix $E_1$. 
Hence, it is not at all obvious that they can be fulfilled for a general 
Drinfeld double and then for any arbitrary choice of the subgroup $H\subset G$.
An obvious simplification occurs when 
$\tilde G$ is an Abelian group.
Then $\Pi^{AB}=0$, $\tilde f^{AB}{}_C=0$ and eq. (\ref{rel1})
is solved by $E_1^{\a\b} \sim \d^{\a\b}$. Then (\ref{actn1}) with 
$\S_{\a\b}\sim \d_{\a\b}$ takes the form of the usual $\s$-model action 
on the coset $G/H$ space. Accordingly \eqn{actn2} represents its usual
non-Abelian dual. Hence, when both groups $G$ and $\tilde G$ are non-Abelian,
the models (\ref{actn1}) and (\ref{actn2}) 
are deformations of the usual 
models on coset spaces $G/H$ and of their non-Abelian duals.

\subsection{ Duality-invariant formulation}

We would like to find a duality-invariant action, from which the $\s$-models 
\eqn{actn1} and \eqn{actn2} originate. It is natural to start 
with the manifestly \PL-invariant action of \cite{DriDou}
from which the $\s$-models \eqn{action1} and \eqn{action2} originate.
This action is defined in the Drinfeld double as \cite{DriDou}, 
\be
 S(l)  =  I_{0}(l) + {1\ov 2\pi} 
\int dx dt \langle{l\inv \del_x l |R| l\inv \del_x l \rangle} \ ,
\label{actiL}
\ee
where $I_{0}(l)$ is the WZW action for a group element $l\in D$.
The operator $R$ is defined as \cite{DriDou}
\be
R = |R^+_A {\rangle} \eta^{AB} \langle{ }R^+_B| 
+ |R^-_A {\rangle} \eta^{AB} \langle{ }R^-_B|  ~ ,
\label{RRpRm}
\ee
with
\be
R^\pm_A  =  T_A \pm (E_0^\pm)_{AB} \tilde T^B ~ ,
~~~~ \eta_{AB}= (E_0^+)_{AB} + (E_0^-)_{AB} ~ ,
\label{rree}
\ee

\no
where we have used the notation $E_0^+= E_0$ and $E_0^- = E_0^T$.
In the limit \eqn{eoin} we have

\be
\pmatrix{R^\pm_a \cr R_\a^\pm } 
\approx  \pmatrix{T_a \cr T_\a \pm (E_1^\pm)\inv_{\a\b} 
\tilde T^\b}\ .
\label{hjf}
\ee

\no
Using this and the conditions \eqn{rel1}, one can show that \eqn{actiL}, 
in the limit \eqn{eoin}, develops the gauge invariance 
\be
l\to l h\ ,\qq  h(t,x)\in H\ ,
\label{sd}
\ee
provided that the following constraint is obeyed
\be
\langle l\inv \del_x l| T_a\rangle = 0\ , \qq \forall\ T_a\ .
\label{coonn}
\ee
In order to avoid introducing this constraint we may use 
gauge fields instead. Indeed, consider the action
\be
 S(l,A_t)  =  I_{0}(l) + {1\ov 2\pi} 
\int \langle{l\inv \del_x l |R_{g/h}|l\inv \del_x l \rangle} 
-{1\ov \pi} \int \langle l\inv \del_x l| A_t\rangle \ ,
\label{gauac}
\ee

\no
where $A_t$ takes values in the Lie algebra of $H$, 
i.e. $A_t = A_t^a T_a$.
The operator $R_{g/h}$ is defined as the restriction in $G/H$ 
of the corresponding operator in \eqn{rree}

\be
R_{g/h} = |R^+_\a  {\rangle} \eta_1^{\a\b} \langle{ }R^+_\b| 
+ |R^-_\a {\rangle} \eta_1^{\a\b} \langle{ }R^-_\b|  ~ ,
\label{RRgh}
\ee
where 
\be
R^\pm_\a  =  T_\a \pm (E_1^\pm)\inv_{\a\b} \tilde T^\b\ ,\qq
(\eta_1)_{\a\b}= (E_1^+)\inv_{\a\b} + (E_1^-)\inv_{\a\b} ~ ,
\label{rrgh}
\ee

\no
and $\eta_1^{\a\b}$ is the inverse matrix of $(\eta_1)_{\a\b}$.
It can be shown that \eqn{gauac} is gauge-invariant under \eqn{sd} and 
the corresponding transformation for the gauge field
\be 
A_t \to h\inv (A_t - \del_t) h \ ,
\label{traa}
\ee
provided that $R_{g/h}$ is invariant under the similarity transformation 
\be 
h R_{g/h} h\inv = R_{g/h} \ .
\label{ghds}
\ee
In order to prove \eqn{ghds} we first show that

\be
h R^\pm_\a h\inv =  \D^\pm_\a{}^\b(h) R^\pm_\b \ ,\qq
\D^\pm_\g{}^\a(h) \D^\pm_\d{}^\b(h) \eta_1^{\g\d} = \eta_1^{\a\b}\ ,
\label{ashjg}
\ee

\no
for some $h$-dependent matrix $\D^\pm_\a{}^\b$.
After repeatedly using \eqn{abpi} and a lengthy computation we find that 
such a matrix exists and is given by 
\be
\D^\pm_\a{}^\b(h) = (E_1^\pm)\inv_{\a\g}(E_1^\pm)^{\d\b} a_\d{}^\g(h) \ ,
\label{dmatr}
\ee
provided that the following condition holds:\footnote{The various algebraic 
manipulations are facilitated by the fact that matrix elements $a(h)_A{}^B$ and
$b(h)^{AB}$ vanish if their indices are not both Greek or Latin.}
\be 
a_{\g}{}^{\a}(h) \a_\d{}^\b(h) (E_1^\pm)^{\g\d} 
= (E_1^\pm)^{\a\b} \pm \Pi^{\a\b}(h) \ ,
\label{kjd}
\ee
or equivalently, splitting into the symmetric and antisymmetric parts, 
\be
\a_\g{}^\a(h) \a_\d{}^\b(h) S^{\g\d} = S^{\a\b} \ ,\qq
\a_\g{}^\a(h) \a_\d{}^\b(h) A^{\g\d} = A^{\a\b} + \Pi^{\a\b}(h)\ .
\label{kjd1}
\ee 
At first sight it seems that \eqn{kjd} is more restrictive than the 
corresponding conditions in \eqn{rel1}, since, unlike \eqn{rel1},
they are valid for finite-gauge transformations.
However, we show in the appendix that \eqn{rel1} actually implies \eqn{kjd}. 

In the remainder of this subsection, we consider the classical equations of
motion for the (manifestly) duality and gauge-invariant action \eqn{gauac}.
Its variation with respect to all fields is 
\be 
\d S(l,A_t) = -{1\ov \pi} \int \d(l\inv \del_x l) 
\left(l\inv \del_t l  - R_{g/h} l\inv \del_x l + A_t\right) + \d A_t 
l\inv \del_x l \ .
\label{eqmo}
\ee

\no
Specializing to subgroup and coset space indices, we find the equations 
of motion
\ba
 \d(l\inv \del_x l) &:& \qq 
\langle l\inv \del_\pm l | R^\mp_\a \rangle = 0 \ ,\quad 
\langle l\inv \del_t l | T_a \rangle = 0\ ,
\nonumber \\
 \d A_t &:& \qq \langle l\inv \del_x l | T_a \rangle = 0\ ,
\label{jgk}
\ea
where we have used also the fact that, because of \eqn{bili}, 
$\langle A_t |T_a\rangle =0$.
Hence, the constraint \eqn{coonn} follows as the equation of motion for $A_t$. 
Using \eqn{hjf}, the equations of motion in \eqn{jgk} can be cast into 
the form $\langle l\inv \del_\pm l|R^\mp_A\rangle =0$. These have the same
form as the equations of motion for the action \eqn{actiL} \cite{DriDou}.

We finally note that the action \eqn{actiL} is manifestly 
invariant under the transformation $l \to l_0(t) l$ 
for some $t$-dependent group element $l_0\in D$ \cite{DriDou}. 
By introducing gauge fields this symmetry can be promoted into a gauge symmetry
with $l_0$ a function of $t$ and $x$.
This type of gauge invariance, though interesting enough in its own right
to be further investigated,
has no apparent relation to the one we have just discussed.

\subsection{The canonical transformation} 

\PL-related models are canonically equivalent under the transformation 
\cite{PLsfe1,PLsfe2}
\ba
&& \tilde P^A = J^A \ ,\qq \tilde J_A = P_A\ ,
\nonumber \\
&& J^A= L^A_x + \Pi^{AB} P_B\ ,\qq 
\tilde J_A= \tL_{x A} + \tilde \Pi_{AB} \tilde P^B  \ .
\label{cabn}
\ea
This transformation preserves the equal-time 
Poisson brackets of the conjugate pairs of 
variables $(J^A,P_A)$ and $(\tilde J_A,\tilde P^A)$ given 
by \cite{PLsfe2}\footnote{We 
will not display explicitly the 2-dim space-time dependence of the phase-space
variables involved in the various Poisson brackets. It is understood that the
first one in the bracket is always evaluated at $x$ and the second one at $y$,
whereas the $t$-dependence is common. Also, compared with \cite{PLsfe2},
we have restored in the various Poisson brackets the
dependence on the scale $\l$.}
\ba
\{J^A,J^B\} & = & \tilde f^{AB}{}_C J^C \d(x\!-\!y) ~ ,
\nonumber \\
\{P_A,P_B\} & = & f_{AB}{}^C P_C \d(x\!-\!y)  ~ ,
\label{JPJP} \\
\{J^A,P_B\} & = & \left( f_{BC}{}^A J^C - \tilde f^{AC}{}_B P_C \right) 
\d(x\!-\!y) + {1\ov \l} \d_B{}^A \d^\prime(x\!-\!y) ~ ,
\nonumber
\ea
and
\ba 
\{\tilde J_A,\tilde J_B\}& =& f_{AB}{}^C \tilde J_C \d(x\!-\!y) ~ ,
\nonumber \\
\{\tilde P^A,\tilde P^B\}& = & \tilde f^{AB}{}_C \tilde P^C \d(x\!-\!y)  ~ ,
\label{tJPJP} \\
\{\tilde J_A,\tilde P^B\}& =&\left( \tilde f^{BC}{}_A \tilde J_C 
- f_{AC}{}^B \tilde P^C \right) \d(x\!-\!y) 
+ {1\ov \l} \d_A{}^B \d^\prime(x\!-\!y) ~ ,
\nonumber
\ea
where $\e(x\!-\!y)$ is the antisymmetric step function that equals $+1$($-1$)
for $x>y$ ($x<0$).
Notice that the above Poisson brackets are independent of the details of the 
$\s$-models related by \PL. They are simply the central extensions, 
in loop space,
of the usual Lie-(bi-)algebras defined in the Drinfeld double.
One may also show that the Hamiltonians of the two dual actions 
\eqn{action1} and \eqn{action2} are equal \cite{PLsfe1} as required for 
canonical transformation with no explicit $t$-dependence.
After some algebraic manipulations, these Hamiltonians can be written as 
\be
H = {\l\ov 2} J^A (G_0-B_0 G_0\inv B_0)_{AB} J^B + {\l\ov 2}
P_A (G_0\inv)^{AB} P_B
- \l J^A(B_0 G_0\inv)_A{}^B P_B \ ,
\label{haml}
\ee
and
\be 
 \tilde H ={\l\ov 2}  \tilde J_A 
(\tilde G_0-\tilde B_0 \tilde G_0\inv \tilde B_0)^{AB} \tilde J_B 
+ {\l\ov 2} \tilde P^A (\tilde G_0\inv)_{AB} \tilde P^B
- \l \tilde J_A (\tilde B_0 \tilde G_0\inv)^A{}_B \tilde P^B \ ,
\label{hamldu}
\ee
where $G_0$ and $B_0$ are the symmetric and antisymmetric parts of $E_0^+$
and similarly $\tilde G_0$ and $\tilde B_0$ are 
the symmetric and antisymmetric parts of $(E_0^+)\inv$.\footnote{The proof 
that $\tilde H=H$ uses the fact that 
\be 
\tilde G_0\inv =  G_0-B_0 G_0\inv B_0\ ,\qq \tilde B_0 \tilde G_0\inv =
- G_0\inv B_0\ ,
\label{gobo}
\ee
as well as the similar 
expressions obtained by interchanging tilded and untilded symbols.
}
Notice that in the limit \eqn{eoin} the conjugate momenta $P_a$ vanish.
This is consistent with the development of a local gauge invariance 
\eqn{trfin}. At the level of the Poisson brackets the vanishing of $P_a$,
together with its conjugate $J^a$, has to be imposed as a constraint.
In fact they form a set $\varphi_a=(P_a,J^a)$ of second-class constraints.
We may see that in the limit \eqn{eoin} and upon using \eqn{rel2},
the Hamiltonians \eqn{haml} and \eqn{hamldu} reduce to
\be
H_{G/H} = {\l\ov 2} (S\inv)_{\a\b} J^\a J^\b + {\l\ov 2} S^{\a\b} P_\a P_\b 
+ \l (S\inv A)_\a{}^\b J^\a P_\b\ ,
\label{amco1}
\ee
and 
\be 
\tilde H_{G/H} = {\l\ov 2} S^{\a\b} \tilde J_\a \tilde J_\b + 
{\l\ov 2} (S\inv)_{\a\b} 
\tilde P^\a \tilde P^\b 
+ \l (S\inv A)_\a{}^\b \tilde P^\a  \tilde J_\b\ .
\label{amco2}
\ee
We may show, with the help of \eqn{JPJP} and \eqn{tJPJP},
that $\{H_{G/H},P_a\}=\{H_{G/H},J^a\}\simeq 0$ (weakly).
Hence, no new constraints are generated by the time $t$-evolution.

In general (see, for instance, \cite{dirac}),
in the presence of a set of second-class 
constraints $\{\varphi_a\}$, one computes the antisymmetric
matrix associated with their Poisson brackets 
$D_{ab}=\{\varphi_a,\varphi_b\}$. When $D_{ab}$ is invertible one simply 
postulates that the usual Poisson brackets are replaced by Dirac brackets, 
defined as 
\be
\{A,B\}_D = \{A,B\}  - \{A,\varphi_a\} (D\inv)^{ab} \{\varphi_b,B\}\ ,
\label{diir}
\ee
for any two phase-space variables $A$ and $B$.
In our case we compute the (infinite-dimensional) matrix
\be
D(x,y) = 
{1\ov \l} \pmatrix{0 & \d_a{}^b\cr \d_a{}^b & 0 } \d^\prime(x\!-\!y) \ ,
\label{dxy}
\ee
with inverse
\be
D\inv(x,y) ={\l\ov 2} \pmatrix{0 & \d_a{}^b\cr \d_a{}^b & 0 } \e(x\!-\!y) \ .
\label{dxyinv}
\ee
Then the Dirac brackets can be computed using \eqn{diir}. We find (for 
notational convenience in the
rest of the paper, we omit the subscript $D$ from the Dirac brackets):
\ba
\{J^\a,J^\b\}&=&\tilde f^{\a\b}{}_\g J^\g \d(x\!-\!y) 
- {\l\ov 2} \e(x\!-\!y) F_1^{\a\b}(x,y)
\ ,\nonumber \\
F_1^{\a\b}(x,y) 
&\equiv & (f_{c\g}{}^\a \tilde f^{c\b}{}_\d + f_{c\d}{}^\b \tilde  
f^{c\a}{}_\g) J^\g(x) J^\d(y)  
\label{jjpa1}\\
&&  - \tilde f^{\a\g}{}_c \tilde f^{c\b}{}_\d P_\g(x) J^\d(y) 
- \tilde f^{\b\g}{}_c \tilde f^{c\a}{}_\d P_\g (x) J^\d(x) \ ,
\nonumber 
\ea
\ba
\{P_\a,P_\b\} & = & f_{\a\b}{}^\g P_\g \d(x\!-\!y) -{\l\ov 2}
\e(x\!-\!y) (F_2)_{\a\b}(x,y)\ ,
\nonumber \\
(F_2)_{\a\b}(x,y) 
&\equiv &(\tilde f^{c\g}{}_\a  f_{c\b}{}^\d + \tilde f^{c\d}{}_\b 
f_{c\a}{}^\g) P_\g(x) P_\d(y) 
\label{jjpa2}\\
&&  - f_{\a\g}{}^c f_{c\b}{}^\d J^\g(x) P_\d(y) 
-  f_{\b\g}{}^c  f_{c\a}{}^\d J^\g (y) P_\d(x) \ ,
\nonumber
\ea
\ba
\{ J^\a,P_\b\} & = & 
\left(f_{\b\g}{}^\a J^\g -\tilde f^{\a\g}{}_\b P_\g\right)\d(x\!-\!y)
+ {1\ov \l} \d^\a{}_\b \d^\prime(x\!-\!y) -{\l\ov 2} \e(x\!-\!y) F^\a_{3\b}(x,y)\ ,
\nonumber \\
F^\a_{3\b}(x,y) &\equiv &  \left( f_{c\g}{}^\a J^\g(x) 
- \tilde f^{\a\g}{}_c P_\g(x)\right) \left( f_{\b\d}{}^c J^\d(y) 
- \tilde f^{c\d}{}_\b P_\d(y) \right) 
\label{jjpa3}\\
&& + \tilde f^{\a c}{}_\g f_{c\b}{}^\d J^\g(x) P_\d(y)\ .
\nonumber
\ea
Notice the parafermionic character of this algebra,\footnote{This is
reminiscent of the parafermionic algebras that appeared \cite{paraa}
in the study of 
classical aspects of exact conformal field theories 
corresponding to gauged WZW models.}
which is encoded in 
the terms containing $\e(x\!-\! y)$. The Dirac brackets for the
pair $(\tilde J_\a,\tilde P^\a)$ are obtained from \eqn{jjpa1}--\eqn{jjpa3}
by replacing untilded symbols by tilded ones and vice versa.
It is instructive to write down the Dirac brackets for the case that the group
$\tilde G$ is Abelian, i.e. $\tilde f^{AB}{}_C=0$.
We find 
\ba
\{J^\a,J^\b\} &=& 0\ ,
\nonumber \\
\{P_\a,P_\b\} &= & f_{\a\b}{}^\g P_\g \d(x\!-\!y) 
\nonumber \\
&& +\ {\l\ov 2} \e(x\!-\!y)
\left( f_{\a\g}{}^c f_{c\b}{}^\d J^\g(x) P_\d(y) 
+ f_{\b\g}{}^c  f_{c\a}{}^\d J^\g (y) P_\d(x) \right) \ ,
\label{jghgh}\\
\{ J^\a,P_\b\} & =&  f_{\b\g}{}^\a J^\g \d(x\!-\!y)
+ {1\ov \l} \d^\a{}_\b \d^\prime(x\!-\!y) -{\l\ov 2} \e(x\!-\!y) f_{c\g}{}^\a 
f_{\b\d}{}^c J^\g(x)  J^\d(y) \ .
\nonumber 
\ea
The above Dirac brackets can also be obtained from the ones in 
\eqn{jjpa1}--\eqn{jjpa3} via a contraction that Abelianizes the group 
$\tilde G$, i.e. $J^\a\to {1\ov \e} J^\a$, $\l\to \e \l$, $\e\to 0$.

\section{An explicit example}
\setcounter{equation}{0}

In this section we explicitly demonstrate many of the general aspects 
developed in section 2, using 3- and 2-dim models related by 
\PL. That includes the explicit construction of the metric and antisymmetric 
tensor fields, of the Dirac-bracket algebra for
canonical equivalence, and 
also of the corresponding generating functional.

\subsection{The Drinfeld double}

Our example will be based on the 6-dim 
Drinfeld double considered in \cite{PLsfe3,toappp,lledo}, which we first 
review by 
following \cite{PLsfe3}.\footnote{Recently, a
classification was made 
of all possible Drinfeld doubles based on the 3-dim real Lie
algebras (Bianchi algebras) \cite{JR}.
It will be interesting to use them for the construction of more examples
that could be useful for the investigation of 
various issues presented in this and the following section.}
It is just the non-compact group $SO(3,1)$ with $G=SU(2)$ and dual 
$\tilde G =E_3 ={\rm solv }(SO(3,1))$ given by the 
Iwasawa decomposition of $SO(3,1)$ \cite{helgason}.
The associated 3-dim algebras 
$su(2)$ and $e_3$ have generators denoted by $\{T_A\}$ and $\{\tilde T^A\}$, 
where $A=1,2,3$.
Leaving aside the details we only present 
the elements that are necessary in this paper. It is convenient to split the
index $A=(3,\a)$, $\a=1,2$. The non-vanishing structure constants 
for the algebras $su(2)$ and $e_3$ are
\be
f_{\a\b}{}^3 = f_{3\a}{}^\b = \e_{\a\b}\ ,\qq \tilde f^{3\a}{}_\b =\d_{\a\b}\ ,
\label{struu}
\ee 
where our normalization is such that $\e_{12}=\d_{11} =1$.
We parametrize the $SU(2)$ group element in terms of
the three Euler angles $\phi$, $\psi$ and $\th$. It is represented by 
the $4 \times 4$ block-diagonal matrix 
\be
g_{SU(2)}= \diag( g, g ) ~ ,
\label{gsu2}
\ee
where 
\be
g = e^{{i\ov 2}\phi \s_3}  e^{{i\ov 2}\th \s_2}  e^{{i\ov 2}\psi \s_3} =
\pmatrix{ \cos {\th\ov 2} e^{{i\ov 2} (\phi +\psi)} & 
\sin {\th\ov 2} e^{{i\ov 2} (\phi -\psi)} \cr
- \sin {\th\ov 2} e^{-{i\ov 2} (\phi -\psi)} & 
\cos {\th\ov 2} e^{-{i\ov 2} (\phi +\psi)} \cr } ~ .
\label{su211}
\ee
Also the group element of $E_3$ is parametrized in terms of three 
variables $y_1$, $y_2$ and $\chi$ and represented by the following 
$4\times 4$ block-diagonal matrix 
\be
\tg_{E_3}= \diag(\tg_+,\tg_-) ~ ,
\label{ge3}
\ee
where
\ba
&& \tg_+ = \pmatrix { e^{+ {\chi \over 2}} & \chi_+ \cr
0 & e^{-{\chi\over 2}} \cr} ~ ,~~~~
\tg_-  = \pmatrix { e^{- {\chi \over 2}} & 0 \cr
\chi_- & e^{+{\chi\over 2}} \cr} ~ ,
\nonumber \\
&& \chi_\pm = \pm e^{-{\chi\ov 2}} (y_1 \mp i y_2) ~ .
\label{ge22}
\ea
The Maurer--Cartan forms in the parametrization
of the $SU(2)$ group element \eqn{su211} are 
\ba
L^1 &  =&  \cos \psi \sin \th d\phi - \sin \psi d\th ~ ,
\nonumber \\
L^2 & = &  \sin \psi \sin \th d\phi + \cos \psi d\th  ~ ,
\label{mausu2} \\
L^3 & = & d\psi + \cos \th d\phi ~ .
\nonumber 
\ea
Similarly, using the parametrization \eqn{ge3} for the $E_3$ group element
we find
\be
\tL_1 =   e^{-\chi }  dy_1 ~ ,~~~~ \tL_2  =  e^{-\chi }  dy_2 ~ ,~~~~
\tL_3  =  d\chi ~ .
\label{maue3}
\ee
The antisymmetric matrices $\Pi$ and $\tilde \Pi$ are 
\be
\Pi = \pmatrix { 0 & -\sin\psi \sin\th & \cos\psi \sin\th \cr
\sin\psi \sin\th & 0 & 1- \cos \th \cr
- \cos \psi \sin \th & \cos \th -1 & 0 \cr } ~ ,
\label{piii}
\ee
and 
\be
\tilde \Pi= 
\pmatrix{0 &  -y_2 e^{-\chi} & y_1 e^{-\chi} \cr
y_2 e^{-\chi}& 0 & 
-\ha \left(1-(1+ y_1^2 + y_2^2) e^{-2 \chi}\right)  \cr
- y_1 e^{-\chi} &
\ha \left(1-(1+ y_1^2 + y_2^2) e^{-2 \chi}\right)  & 0 \cr } ~ .
\label{ttpiij}
\ee

\subsection{Explicit three- and two-dimensional models}

Consider the $\s$-model action \eqn{action1} for the case of our double based 
on $SO(3,1)$. 
Let us single-out 
the 1-dim subgroup $H\simeq U(1)$ that is generated by $T_3$. 
For our purposes it will be sufficient to use the following 
form for the $3\times 3$ matrix $E_0\inv$
\be
E_0\inv = \pmatrix{(1+g)^{-1} a & 0 & 0\cr
                   0 & a & b-1 \cr
                   0 & 1-b & a}\ ,
\label{eoii}
\ee
where we have kept the conventions of \eqn{bllo} for the enumeration 
of the matrix elements.
Using \eqn{mausu2}, \eqn{piii} and \eqn{eoii}, it is the easy to compute 
the metric and antisymmetric tensor fields 
corresponding to \eqn{action1}. We find a metric given by 
\be
ds^2 = {a\ov V}
\left(
(L^1)^2 +(L^2)^2 +(g+1)(L^3)^2 + {1+ g\ov a^2} \big( (b \cos\th -1)\ d\phi
+(b-\cos\th)\ d\psi\big)^2 \right) \ ,
\label{meett}
\ee
and an antisymmetric tensor given by
\be
B= 2{ \sin\th \ov V}\ d\th\wedge \Big((g+1) d\psi + 
(b + g \cos\th) d\phi\Big)\ ,
\label{btht}
\ee
where 
\be
V \equiv  a^2 + (b-\cos\th)^2 + (1+g)\sin^2\th\ .
\label{veo}
\ee
Notice that the antisymmetric tensor can be (locally) 
gauged away since the corresponding 
3-form field strength is zero.
Also, for our purposes, we will not need the explicit 
expressions for the metric and antisymmetric tensor corresponding to 
the dual $\s$-model \eqn{action2}.
For $b=1$, but general $a$ and $g$,
the above example (with its dual) was considered in \cite{PLsfe3}
(also in \cite{toappp} for $a=b=1$ and $g=0$). 

We would like to take the analogue of the limit \eqn{eoin}. It is clear that
in our case this 
corresponds to letting $g\to -1$. Comparing \eqn{eoii} to \eqn{bllo}
we see that the $2\times 2$ matrix $E_1$ is 
\be
E_1 = \pmatrix{a & b-1 \cr 1-b & a }\ .
\label{e11}
\ee
It is easily seen that this is the 
most general $2\times 2$ matrix 
that solves \eqn{rel1}, with structure constants given by \eqn{struu}.
In agreement with our general discussion, the $\s$-model action with metric 
\eqn{meett} and antisymmetric tensor \eqn{btht} develops a local invariance 
under the transformation
\be
\d \psi=\e(t,x)\ .
\label{trg1}
\ee
This allows to gauge-fix the variable $\psi =0$. 
Explicitly computing \eqn{actn1} we find that the metric and 
antisymmetric tensors are given by 
\ba
ds^2 & = &{a \ov a^2 +(b - \cos\th)^2 } 
\ \left(d\th^2 + \sin^2\th d\phi^2\right) \ ,
\nonumber \\
B & = & 2 {\sin\th (b- \cos\th) \ov  
a^2 +(b-\cos\th)^2 }\ d\th \wedge d\phi \ .
\label{dsbth}
\ea
Equivalently, the same result follows if we set $g=-1$ directly into the 
expressions for the metric \eqn{meett} and antisymmetric \eqn{btht} tensors.
Similarly, the dual model action \eqn{actn2}
is invariant under the local transformation 
\be
\d y_\a = \e(t,x) \e_{\a\b} y_\b\ .
\label{trg2}
\ee
Hence, we may 
evaluate \eqn{actn2} in the gauge $y_1=0$.
The corresponding metric 
(the antisymmetric tensor turns out to be zero) is found to be 
\ba
&& ds^2 = {a_1/2 \ov 1+ a_1 z} \left( {dz^2\ov \r^2} + \left( d\r + \Bigg[
{b-1\ov a} + {z-a_1 \r^2/4\ov 1+ a_1 z}\Bigg] {dz\ov \r} \right)^2 \right)\ ,
\nonumber \\
&& a_1 \equiv {2 a\ov a^2 + (b-1)^2} \ ,
\label{dsbth1}
\ea
where we have changed variables as $y_2^2 = {1\ov 4} \r^2 a_1^2 $ 
and $e^{2 \chi}= 1+  a_1 z$. 
The metric\footnote{If we set $b=1$ and redefine $a\to 2/a $ and $\l\to \l a/2$
the metric \eqn{dsbth} and its dual \eqn{dsbth1} become those 
derived in \cite{PLsfe3} using a limiting procedure,
equivalent to \eqn{eoin}.
The deeper reason that validates such a procedure is, as we have shown 
in the present paper, 
the development of a local gauge invariance in this limit.}
in \eqn{dsbth} is free of singularities (since \eqn{trg1}
has no fixed point) and represents
a deformed 2-sphere. In contrast, \eqn{dsbth1} is singular for 
$r=0$. This is related to the fact that $y_1=y_2=0$ is a fixed point of the 
gauge transformation \eqn{trg2}. 
The singularity at $1+a_1 z=0$ is only a coordinate 
singularity and can be removed by an appropriate change of variables. 

It is worth while 
to consider some analytic continuations of the models \eqn{dsbth}
and its dual \eqn{dsbth1}. If we let $\th\to i r$, 
where $r\in [0,\infty)$, and also we change the sign of the overall coupling
constant $\l$, then \eqn{dsbth} becomes
\ba
ds^2 & = &{a \ov a^2 +(b - \cosh r)^2 } 
\ \left(dr^2 + \sinh^2 r d\phi^2\right) \ ,
\nonumber \\
B & = &  2 {\sinh r (b - \cosh r) \ov  
a^2 +(b-\cosh r)^2 }\ d r \wedge d\phi \ .
\label{dsban}
\ea
The corresponding analytic continuation in the dual metric \eqn{dsbth1}
should be $\r\to i \r$, with a parallel change of sign in the overall 
coupling constant.
The metric in \eqn{dsban} is reduced to the Euclidean $AdS_2$ metric 
if we rescale the coupling constant $\l\to \l/a$ and then 
take the limit $a\to \infty$ (keeping the new coupling finite). 
However, for generic values of the constant
$a$, it represents a space that is topologically a cigar. 
Indeed, for $r\to 0$ we get the 2-dim Euclidean space $E^2$ in polar 
coordinates, whereas
for $r\to \infty$ we get, after an appropriate change of variables, 
$R^1 \times S^1$. For $b>0$, the cigar-shaped space develops a ``pump'' 
corresponding to the 
maximum of the metric components $G_{\phi\phi}$ at 
\be
\cosh r= {\sqrt{(1+a^2+b^2)^2-4 b^2} + 1+a^2+b^2\ov 2 b}\ .
\label{hgsk}
\ee
We note that the 
cigar-shape topology is also a characteristic of the Euclidean black hole 
corresponding to the coset $SL(2,\IR)/U(1)$ exact conformal field
theory \cite{witbla}. 
However, in our case the model \eqn{dsban} is not conformal.
The Drinfeld double for \eqn{dsban} and its dual model 
is $SO(2,2)$, with $G=SL(2,\IR)$, instead of $SU(2)$.

\subsection{The Dirac brackets and the generating functional} 

The Dirac brackets for the conjugate variables in our example are most 
easily written down in the basis $J^\pm = J^1\pm i J^2$ and 
$P_\pm = P_1\pm i P_2$, where the non-zero structure constants are 
$f_{3\pm}{}^\pm=\pm$, $f_{+-}{}^3 =2$ and $\tilde f^{3\pm}{}_\pm =1$.
Using \eqn{jjpa1}--\eqn{jjpa3} we obtain 
\ba
\{ J^\pm, J^\pm\} & = & \mp \l \e(x\!-\!y) 
\underline{{ J^\pm(x) J^\pm(y)} } \ ,
\nonumber \\
\{ J^+,J^-\} & = & 0\ ,
\label{jjss1}
\ea
\ba
\{ P_\pm,P_\pm \} & = & + \l \e(x\!-\!y) \left( \mp 
\underline{ {P_\pm(x) P_\pm(y)}} + 
J^\mp(x) P_\pm(y) + P_\pm(x) J^\mp (y) \right)\ ,
\nonumber \\
\{P_+,P_-\} & = & -\l \e(x\!-\!y) 
\left( J^-(x) P_-(y) + P_+(x) J^+(y) \right)\ ,
\label{jjss2}
\ea
\ba
\{ J^\pm,P_\pm\} & = & {1\ov \l} \d^\prime(x\!-\! y) -\l \e(x\!-\! y)
\left( J^\pm(x) J^\mp(y) \mp \underline{{J^\pm(x) P_\pm(y)}} \right) \ ,
\nonumber \\
\{ J^\pm,P_\mp\} & = & \l \e(x\!-\! y) \left( J^\pm(x) J^\pm(y) 
\pm \underline{{J^\pm(x) P_\mp(y)}} \right ) \ ,
\label{jjss3}
\ea
where the underlined terms should be omitted in the Abelian 
limit of the dual group $\tilde G=E_3$. In this case the above algebra
provides a canonical equivalence between the $\s$-model for 
$S^2$ and its non-Abelian dual
with respect to the left (or right) action of $SU(2)$.
Note also that the generators $J^\pm$ form a subalgebra \eqn{jjss1}.


The generating functional that demonstrates the classical equivalence 
between $\s$-models related by \PL\ based on our Drinfeld double 
was explicitly constructed in \cite{PLsfe3}. 
In a slightly different form than that in \cite{PLsfe3}, it 
reads\footnote{We also correct a misprint in 
eq. (26) of \cite{PLsfe3}. In the expression for $B_\psi$ and in the
argument for $\cot\inv$, $(y_1 \cos\psi +y_2 \sin\psi)$ should be replaced
by $(y_1 \cos\psi +y_2 \sin\psi)\tan{\th\ov 2}$.} 
\ba
F & =& \int dx  \Big( A \del_x \phi + (\psi +\a -\phi)\del_x \chi 
-{2 \rho \tan\inv B \ov \sqrt{1+\rho^2 \cos^2(\psi+\a)}} \del_x
(\rho \cos(\psi+\a)) \Big) \ ,
\nonumber\\
A & \equiv & - \ln \left( e^{2\chi} \cos^2{\th\ov 2} + e^\chi \rho \sin\th
\sin(\psi+\a) + (1+\r^2) \sin^2{\th\ov 2} \right)\ ,
\label{fab}\\
B &\equiv & {e^\chi \cot {\th\ov 2}  + \r \sin(\psi+\a) \ov 
\sqrt{1+\rho^2 \cos^2(\psi+\a)} }\ ,\qq (y_1,y_2)=\r (\sin\a,\cos\a)\ .
\nonumber 
\ea
Notice that the above 
generating functional depends only on the combination
$\psi +\a$; it is therefore invariant under the $U(1)$ gauge 
transformation $\d \psi = \e$ and $\d \a = -\e$. The generating functional 
for the deformed coset models \eqn{dsbth} is obtained by solving 
the equation ${\d F \ov \d \psi} =0$ (equivalently ${\d F \ov \d \a} =0$)
for $\psi +\a$ and inserting the result back 
into \eqn{fab}.\footnote{Such a procedure is motivated by the fact that 
the variations ${\d F \ov \d \psi}$ and
$-{\d F \ov \d \a}$, corresponding to the conjugate momenta $P_\psi$ and 
$P_\a$, are zero since the variables $\psi$ and $\a$ have dropped out of the
corresponding dual $\s$-models because of the gauge invariance.
Also, thanks to the latter, only one of these variations is independent.
This procedure has an obvious generalization for the more general 
coset models constructed in section 2.
For some similar considerations, see \cite{ouggroi2} and more recently 
\cite{stern}.}
The result is a generating functional, which is
non-polynomial in derivatives with respect to $x$. The obtained 
expressions are quite complicated and not very illustrative, so that we
decided to present the corresponding result for the $\s$-model for 
$S^2$ and its non-Abelian dual.
We start with the generating functional corresponding to the 2-dim $\s$-models 
for $S^3$ and its non-Abelian 
dual with respect to the left (or right) action of $SU(2)$ that was obtained
in \cite{zacloz}. In our notation it is given by $F=-\int dx (y_1 L^1_x+
y_2 L^2_x + z L^3_x)$. This is easily modified to depend on the 
angles $\psi$
and $\a$ only through the combination $\psi+\a$, by adding the term 
$-\int dx \a\del_x z$. Such a term, being dependent on the variables
of only one of the dual models, can be absorbed as total derivative into
the corresponding action and hence it does not affect the classical dynamics.
Explicitly, the resulting generating functional is 
\be
F = - \int dx \Big((z \cos \th + \r \sin\th \sin (\psi +\a) )\del_x \phi
+ \r \cos (\psi+\a) \del_x \th - (\psi+\a) \del_x z \Big) \ .
\label{fgh}
\ee
The variation of $F$ with respect to $\psi+\a$ gives 
\be 
\tan (\psi+\a) = { \del_x z \sqrt{ \r^2\left((\del_x\th)^2 
+ \sin^2\th (\del_x \phi)^2\right) - (\del_x z)^2} - \r^2\sin\th \del_x\th
\del_x \phi 
\ov
(\del_x z)^2 - \r^2 (\del_x \th)^2}\ . 
\label{phisa}
\ee
Substituting back into \eqn{fgh} we obtain 
\be
F = - \int dx \left(\sqrt{\r^2\left( (\del_x \th)^2 +
\sin^2 \th (\del_x \phi)^2 \right)- (\del_x z)^2} 
+ z \cos\th \del_x \phi -  (\psi + \a) \del_x z\right) \ ,
\label{fgh1}
\ee
where $\psi+\a$ is given by \eqn{phisa}. 
The generating functional \eqn{fgh1} is non-polynomial in 
the derivatives of the fields with respect to $x$.
In that sense it belongs to a new class of 
generating functionals, which depend 
not only on the fields of the two dual 
$\s$-models, but also on their first  
derivatives with respect to the space-like variable in a non-trivial 
way.
For comparison, up to now, either in the case of non-Abelian duality 
\cite{zacloz,loz,sfepara} or for \PL\ (and its possible generalizations) 
\cite{PLsfe1,PLsfe2}, 
there was no dependence of the generating functional on more than the
first power of these derivatives (see, for example \eqn{fab}).\footnote{A
generating functional of the type \eqn{fgh1}, containing first derivatives
of the fields in a non-polynomial way, has appeared in a 
study on the canonical equivalence between Liouville and free field 
theories \cite{FS} (also \cite{FS1} as quoted in \cite{FS}).}
Finally, we note that, according to the work in \cite{qufun}, 
generating functionals of the form \eqn{fgh1}, being non-linear, are 
expected to receive quantum corrections. Consequently, the corresponding 
duality rules relating the 2-dim field theories, as well as the algebra
\eqn{jjpa1}--\eqn{jjpa3}, 
are expected to be quantum-corrected.

\section{Renormalization group flow}
\setcounter{equation}{0}

In this section we study the 1-loop RG equations 
corresponding to the three-dimensional model \eqn{meett}, \eqn{btht}.
We will show that there are no fixed points in the flow and also that 
the correct description of the models is a non-perturbative one. 
However, for large domains in parameter space 
and for a wide range of energies in the UV, the description 
is effectively perturbative and the model becomes a two-dimensional one.
Finally, by performing some analytic continuations we will 
find three- and two-dimensional models with fixed points under the RG flow,
where the theory becomes free.

We begin this section with a short review of RG flow in 
2-dim field theories with curved target spaces.
and \eqn{dsbth}. 
The 2-dim $\s$-model corresponding to the metric \eqn{meett} and 
antisymmetric tensor \eqn{btht} is of the form 
\be 
S={1\ov 2\l} \int Q^+_{\m\n} \del_+ X^\m \del_- X^\n\ ,\qq 
Q^+_{\m\n}\equiv G_{\m\n} + B_{\m\n}\ .
\label{fjll}
\ee  
It will be renormalizable if the corresponding counter-terms, at
a given order in a loop expansion, 
can be absorbed into a renormalization of the 
coupling constant $\l$ and (or) of 
some parameters labelled collectively by $a^i$, $i=1,2,\dots$
In addition, we allow for general field redefinitions of the $X^\m$'s, 
which are coordinate reparametrizations in the target space.
This definition of renormalizability of $\s$-models is quite strict and
similar to that for ordinary field theories. A natural extension of this is to
allow for the manifold to vary with the mass scale and the RG
to act in the infinite-dimensional space of all metrics and 
torsions \cite{Frialv}. 
Further discussion of this generalized renormalizability will
not be needed for our purposes.
Perturbatively, in powers of $\l$, we express the bare quantities, denoted 
by a zero as a subscript, as
\ba
&&\l_0 = \m^\e \l \left(1+ {J_1(a)\ov \pi \e} \l +
\cdots \right) \equiv \m^\e \l
\left (1+ {y_\l \ov \e} + \cdots \right )  ~ ,
\nonumber \\
&&a^i_0 = a^i + {a^i_1(a) \ov \pi \e} \l  
+ \cdots  \equiv a^i 
\left (1+ {y_{a^i} \ov \e} + \cdots \right )  ~ ,
\label{expp}\\
&&X^\m_0 = X^\m + {X^\m_1(X,a) \ov \pi \e} \l + \cdots  ~ .
\nonumber
\ea
The ellipses stand for higher-order loop- and pole-terms in $\l$ and
$\e$ respectively.
Then, the beta-functions up to one loop are given by 
$\b_\l = \l^2 {\del y_\l\ov \del \l}
= {\l^2\ov \pi} J_1$ and 
$ \b_{a^i} = \l a^i {\del y_{a^i}\ov \del \l}=
{\l \ov \pi} a^i_1$, where, as usual, $\b_\l = {d\l\ov dt}$, 
$\b_{a^i} = {da^i\ov dt}$ and $t=\ln \m$.
The equations to be satisfied by appropriately choosing $J_1,a^i_1$ 
and $X_1^\m$ are given by
\be
\ha R^-_{\m\n} = 
-J_1 Q^+_{\m\n} + \del_{a^i} Q^+_{\m\n} a^i_1 + \del_\l Q^+_{\m\n} X_1^\l
+  Q^+_{\l\n} \del_\m  X_1^\l  +  Q^+_{\m\l} \del_\n X_1^\l\ , 
\label{1loop}
\ee
where $R^-_{\m\n}$ are the components of 
the ``generalized'' Ricci tensor defined with a connection 
that includes the torsion, i.e. with $\G^{\m}_{\n\r} - \ha H^\m{}_{\n\r}$.
The corresponding counter-terms were computed in the dimensional
regularization scheme (see, for instance, \cite{Osborn}).

\subsection{Models with no fixed points}

\subsubsection{Three-dimensional models}

In the metric \eqn{meett} there are three parameters $a$, $b$ and $g$ and
the three Euler angles $\th,\psi$ and $\phi$ will be denoted by $X^\m$.
Also for the antisymmetric tensor in \eqn{btht} we have $H^\m{}_{\n\r}=0$.
Examining \eqn{1loop} we find that the coupling $\l$ 
and the coordinates $(\th,\psi,\phi)$
do not renormalize and therefore the corresponding beta-functions are
zero.\footnote{We believe that the non-renormalization of the
overall coupling constant $\l$ will persist in general for all 
\PL-related models with actions \eqn{action1} and \eqn{action2}. 
On the other hand, models corresponding 
to a limit of \eqn{meett} and with target space $S^3$ or its deformation 
along a
direction in the Cartan subalgebra of $SU(2)$ \cite{ouggroi} (see also 
the comments after \eqn{abgli} below), have an overall
coupling constant that gets renormalized \cite{ouggroi}. 
The reason for this apparent paradox
is that, in these models, the overall coupling constant is related to our $\l$
by rescalings, such as those described in footnote 8, with parameters that 
get renormalized.}
In contrast, for the parameters $a,b$ and $g$ we find
\ba
\b_a& = & {\l\ov 4\pi} {1+a^2-b^2\ov a^2} 
\Big((g-1) a^2 + (g+1) (b^2-1)\Big)\ ,
\nonumber\\
\b_b & = & {\l\ov 2\pi} {b\ov a}  \Big((g-1) a^2 + (g+1) (b^2-1)\Big)\ ,
\label{abg1}\\
\b_g &=& {\l\ov 2\pi} {1+g\ov a}  \Big(g (1+a^2) + (g+2) b^2\Big)\ .
\nonumber 
\ea
This system of coupled non-linear equations\footnote{Presumably, 
the dual to the \eqn{meett}, \eqn{btht} model will also 
have the same beta-functions 
\eqn{abg1}.
We also note that it is highly non-trivial that the change of the matrix 
\eqn{eoii} under the RG eqs. \eqn{abg1} preserves its
form. For example, had we started, as in \cite{toappp},
with a matrix $E_0\inv $ proportional to the identity, then eqs.
\eqn{abg1} would have generated off-diagonal elements.
This is clearly seen by computing the right-hand sides of 
eqs. \eqn{abg1}
for $a=b=1$ and $g=0$. By doing so we obtain the infinitesimal change 
of $E_0\inv =I$ (to lowest order) and eq. (102) of \cite{toappp} 
(with $c=1$).}
can be considerably simplified.
First, using \eqn{abg1}, we may easily show that 
there is a RG-flow-invariant defined as 
\be
{a^2+b^2+1\ov b} \equiv 2 \n  = {\rm const.}\ ,
\label{invv}
\ee
which implies that 
\be
a=\sqrt{(b_+ - b)(b-b_-)}\geq 0\ ,\qq b_\pm\equiv \n\pm \sqrt{\n^2-1}\ ,
\qq |\n|\geq 1 \ .
\label{hja}
\ee
%
%
Without loss of generality we may assume that $\n>0$ since \eqn{invv} 
remains invariant under $\n\to -\n$ and $b\to -b$.
Then, using the last two equations 
in \eqn{abg1} we may derive an equation for $b$ as
a function of $g$ whose solution is 
\be
b=- g\left(\n \pm \sqrt{\n^2 -1+e^{-2 C} (1+1/g)^2}\ \right)\ ,
\label{bgy}
\ee
where $C$ is a real constant, which is determined by the initial conditions
for $b$ and $g$. The sign in front of the square root in \eqn{bgy} 
is changed when $g=0$, in order to ensure the continuity of $b$ as a function 
of the energy scale $t=\ln \m$.
%
Hence, the only differential equation we still have to solve is the one 
for $g$, which, after using \eqn{invv}, takes the form
\be
\b_g =  {\l\ov \pi} {b \ov a}\ (g+1) (b+\n g)\ ,
\label{abg11}
\ee
where $a$ and $b$ are determined by \eqn{hja} and \eqn{bgy}.
Since the RG equations are real, $a^2$ will stay strictly 
non-negative and therefore $b$ will oscillate with $t=\ln \mu$
between its minimum and maximum values $b_-$ and $b_+$, where $a=0$. 
When $a\simeq 0$, for finite values of the overall coupling constant $\l$, 
the curvature for the metric \eqn{meett} approaches infinity and the 
perturbative expansion of the RG
equations becomes meaningless.

We have seen that the correct description of the theory is 
a genuine non-perturbative one. Neverthelss, for $\n\gg 1$ we will 
show that there exists a 
wide range of energies in the UV, where the description is effectively 
perturbative. Moreover, there exists a fixed point at $g=-1$ where
the theory has effectively a 2-dim target space.
Indeed, using \eqn{invv}, we have that 
$a^2\simeq 2 \n b\gg 1$ when $\n \gg 1$. Hence, in that limit and after 
redefining $\l\to \l/a$ we may simplify the RG eqs. \eqn{abg1} as
\ba
\b_\l& \simeq & -{\l^2\ov 4\pi} (1-g)\ ,
\nonumber\\
\b_g &\simeq & {\l\ov 2\pi}\ g (1+g) \ ,
\label{abgli}\\
\b_b & \simeq &- {\l\ov 2\pi}\ (1-g) b \ .
\nonumber 
\ea
Then the metric \eqn{meett} becomes 
\ba
ds^2 & =&  (L^1)^2 + (L^2)^2 + (1+g) (L^3)^2
\nonumber \\
& = & d\th^2 + \sin^2\th d\phi^2 + (1+g) (d\psi + \cos\th d\phi)^2\ ,
\label{jef}
\ea
which is the deformed $SU(2)$ Principal
Chiral model considered in \cite{ouggroi}. Also 
the first two of the above equations are those derived in \cite{ouggroi}
for the corresponding coupling $\l$ and deformation parameter $g$.
In the UV the solution of \eqn{abgli} is
\be 
\l \simeq {2\pi \ov t} ,\qq g \simeq -1 +{{\rm const.}\ov t}\ ,
\qq b\simeq {{\rm const.}\ov t^2}
\ ,\qq {\rm as} \ t\to \infty \ .
\label{asdj}
\ee
Hence, in the UV $a^2 \simeq 2 \n b \sim 2\n/t^2$. 
Therefore if the condition
\be
1 \ll t \ll \n^{1/2}\ ,
\label{sdsf}
\ee
is fulfilled, then $a\gg 1 $ and the model is indeed 
described perturbatively by \eqn{jef}. The point $g=-1$ is a UV-fixed point,
where the metric \eqn{jef} becomes $S^2$.
However, outside the validity of \eqn{sdsf} 
the correct description is non-perturbative.

\subsubsection{Two-dimensional models}

Let us now return to the 2-dim models \eqn{dsbth} and \eqn{dsbth1}.
As before, there is no wave-function renormalization for $\th$ and $\phi$,
and the beta-function for the coupling $\l$ is zero. For the couplings $a$ 
and $b$ the corresponding beta-functions can be obtained by simply
setting $g=-1$ into \eqn{abg1}. The reason why such a procedure is 
consistent seems to be intimately related to the local invariance 
that reduces the 3-dim models into 2-dim ones. Hence, we have
\ba
\b_a& = & - {\l\ov 2\pi}\ (1+a^2-b^2)\ ,
\nonumber\\
\b_b & = & - {\l\ov \pi}\ ab\ ,
\label{abg2}
\ea
which are nothing but the beta-functions for the 2-dim 
model \eqn{dsbth} as well as for its dual \eqn{dsbth1}.\footnote{In order 
to compare with $\b_a$ and $\b_\l$
as given by eq. (47) of \cite{PLsfe3}, one should remember that these 
correspond to the model \eqn{dsban} with $b=1$.
Imposing that $b=1$
and further requiring that $\b_b=0$ enforces 
a wave-function renormalization of the $\th$, i.e. in \eqn{expp}
we have $\th_1 =-{1\ov a} \sin\th$,
in order for the model to be 1-loop-renormalizable. Then it turns out that
$\b_a= -{\l\ov 2\pi}(4+a^2)$.  
After taking into account the redefinitions of the various parameters, as
described in footnote 8 of the present paper, this 
implies eq. (47) of \cite{PLsfe3}.}
This is a strong hint that their classical equivalence can be 
promoted into a quantum one as well. Having said that we note, once again,  
that $g=-1$ is not a fixed point of the \eqn{abg11} in the UV.
Since \eqn{invv} is still a RG invariant of \eqn{abg2},
it is clear that one variable between $a$ and $b$ is an independent one.
Eliminating $a$ from \eqn{abg2} using \eqn{invv}, we obtain 
\be
\b_b = -{\l\ov \pi} b \sqrt{(b_+-b)(b-b_-)}\ .
\label{ab11}
\ee
Hence, the solution 
for $b$ as a function of the energy scale $t=\ln \m$ oscillates between 
$b_+$ and $b_-$ as
\be 
{1\ov b(t)} = \n + \sqrt{\n^2 -1} \sin {\l \ov \pi} (t-t_0)\ ,
\label{osscc}
\ee
where $t_0$ is an arbitrary reference scale.
This means that the corresponding $\s$-model actions do not
define local field theories and can be considered at most as effective actions
for scales such that $b$ stays away from $b_\pm$.

The usual $S^2$ metric and its non-Abelian dual with respect to the 
right (or left) action of $SU(2)$ are obtained from \eqn{dsbth} and 
\eqn{dsbth1} if we rescale the coupling constant $\l\to \l/a$ and then 
take the limit $a\to \infty$ (keeping the new coupling finite). 
However, this limit is
problematic at the quantum level since the corresponding $\b$-functions
do not tend to the beta-function obtained by studying the 2-dim field 
theories based on $S^2$ (and its non-Abelian dual) by themselves \cite{PLsfe3}.
The latter is, at one-loop, just $\b_\l = -{\l^2\ov 2\pi}$ and 
is consistent with the
fact that these models are asymptotically free. It is formally obtained
by the first of \eqn{abg2} in the limit $a\to \infty$ after we rescale $\l
\to \l/a$ as described above. This limit does not correspond to any
fixed point of \eqn{abg2}.
It is easily seen that, from a RG theory view point, 
these models offer 
an effective description of the more general models \eqn{dsbth} and 
\eqn{dsbth1} in the case of 
$\a\simeq b\simeq \n\gg 1$, which, according to \eqn{osscc}, occurs at
scales ${\l\ov \pi} (t-t_0)\simeq -{\pi \ov 2} + {1\ov \n}\ {\rm mod}(2\pi)$.

\subsection{Models with fixed points}

\subsubsection{Three-dimensional models}

We have seen that our model \eqn{meett}, \eqn{btht} does not have a true
fixed point
under the 1-loop RG eqs. \eqn{abg11}. Consider, however,
the analytic continuation $\l\to -i \l$ and $a\to i a$. Then 
the metric and antisymmetric tensors become
\be
ds^2 = {a\ov V}
\left(
(L^1)^2 +(L^2)^2 +(g+1)(L^3)^2 - {1+ g\ov a^2} \big( (b \cos\th -1)\ d\phi
+(b-\cos\th)\ d\psi\big)^2 \right) \ ,
\label{meett1}
\ee
and
\be
B= 2 i {\sin\th \ov V}\ d\th\wedge \Big((g+1) \wedge d\psi + 
(b + g \cos\th) \wedge d\phi\Big)\ ,
\label{btht1}
\ee
where instead of \eqn{veo} the function $V$ is given by 
\be
V \equiv  a^2 - (b-\cos\th)^2 - (1+g)\sin^2\th\ .
\label{veo1}
\ee
The fact that the antisymmetric tensor is imaginary is bothersome if we want
to describe models in 2-dim Minkowskian space-times. However, for Euclidean 
ones, the (locally) exact 2-form measures the charge of 
non-trivial instanton-like configurations.
The perturbative expansion is completely independent of the antisymmetric 
tensor, but this will definitely play a r\^ole in a, yet lacking,
non-perturbative formulation of the model. 
The 1-loop RG equations for the metric \eqn{meett1} 
are obtained from \eqn{abg1} by the analytic continuation 
we have described above. Then the analogue of \eqn{abg11} is given by 
\be
\b_g = - {\l\ov \pi} {b \ov a}\ (g+1) (b+\n g)\ ,
\label{abg111}
\ee
where now
\be
a=\sqrt{(b - b_+)(b-b_-)}\geq 0\ ,\qq b_\pm\equiv \n\pm \sqrt{\n^2-1}\ ,
\label{hja1}
\ee
and $b$ is still given by \eqn{bgy}. As before, we will assume that $\n>0$
with no loss of generality. 
However, now $\n$ does not have to be larger than
or equal to 1, as in \eqn{hja}, in
order to ensure reality for $a$. If $\n<1$ then $b_\pm$ are complex 
conjugate of each other and, unlike the case when they are real, $b$ can take 
any real value without spoiling the reality of the parameter $a$.
However, now the condition $|1+1/g|\geq e^C \sqrt{1-\n^2}$ has to be fulfilled 
in order for $b$ to remain real.
If on the other hand $\n>1$, then $b_\pm$ are both real and the
reality condition for $a$ requires that $b\geq b_+>b_-$ or $b\leq b_-<b_+$.
Since $0<b_-<b_+$, it turns out that  
there are fixed points for initial conditions where $b$ is less than
$b_-$.
Consider first the RG eq. \eqn{abg111} 
near the point with $g=1/(e^C-1)$, $b=0$ and $a=1$. It can be written as
(we take the lower sign in \eqn{bgy}):
\be
\b_g \simeq {\l\ov \pi} g^* (g-g^*)\ ,\qq g^* \equiv 1/(e^C-1)\ .
\label{bg1}
\ee
The same equation near the different 
point with $g=-1/(e^C+1)$, $b=0$ and $a=1$ takes the form
\be
\b_g \simeq {\l\ov \pi} \tilde g^* (g-\tilde g^*)\ ,\qq  \tilde 
g^*\equiv -1/(e^C+1)\ .
\label{bg2}
\ee
For $e^C>1$ we have $-\ha < \tilde g^* <0< g^* $.
Hence, for $e^C>1$ we have an IR-stable point at $g=g^*$ as well as
a UV-stable point at $g=\tilde g^*$. 
For $0<e^C<1$, we have that $g^*<-1<\tilde g^*<-\ha$.
Therefore, for $0<e^C<1$ there are two UV-stable 
points at $g=g^* $ and at $g=\tilde g^*$. 

In all cases the background \eqn{meett1}, \eqn{btht1}
flows, either in the IR or in the UV, towards the background with
\ba
ds^2& = & -{1\ov g_0} \left(  {d\th^2 \ov \sin^2\th } - g_0 d\phi^2
+(g_0+1) d\psi^2 \right)\ ,
\nonumber\\
B& = & -{2 i \ov g_0} {1\ov \sin\th} d\th \wedge \Big((g_0+1) 
d\psi + g_0 \cos\th d\phi \Big)\ ,
\label{ddsf}
\ea
where $g_0$ represents any of the two fixed points $g^*$ or $\tilde g^*$. 
This represents a free theory, as can be seen by changing variables 
as $\sin\th = {1\ov \cosh y}$.
It is interesting to note that in the case $e^C>1$ the 
signature of the metric in \eqn{ddsf} is $(-+-)$ in the IR fixed point 
$g_0=g^*$ and $(+++)$ in the UV fixed point $g_0=\tilde g^*$.
Also in the case of $0<e^C<1$ the signature at the $g_0=g^*$
UV-stable point is $(++-)$, but in the other UV-stable point at $g_0=\tilde 
g^*$ it is $(+++)$.
Hence, only at $g=\tilde g^*$ the metric has Euclidean signature and we expect 
a well-defined field-theoretical description.

Let us also note that for $\n\gg 1$ the RG flow is described, as before, 
by \eqn{abgli}, \eqn{asdj} and the corresponding $\s$-model is again 
\eqn{jef}, provided \eqn{sdsf} is satisfied.

\subsubsection{Two-dimensional models}

Now we turn to the 2-dim model \eqn{dsbth} after the same 
analytic continuation as before, $a\to i a$ and $\l\to -i \l$:
\ba
ds^2 & = &{a \ov a^2 -(b - \cos\th)^2 } 
\ \left(d\th^2 + \sin^2\th d\phi^2\right) \ ,
\nonumber \\
B & = & 2 i {\sin\th (b- \cos\th) \ov  
a^2 -(b-\cos\th)^2 }\ d\th \wedge d\phi \ .
\label{dsbbh}
\ea
The 1-loop RG equation corresponding to \eqn{ab11} is
\be
\b_b = -{\l\ov \pi} b \sqrt{(b-b_+)(b-b_-)}\ .
\label{ac11}
\ee
The form of the solution 
for $b$ as a function of the energy scale $t=\ln \m$ depends on whether 
or not $\n$ is smaller or larger than 1. We find
\be 
{1\ov b(t)} = \n + \sqrt{1-\n^2} \sinh{\l \ov \pi} (t-t_0)\ ,\quad
{\rm if }\quad \n<1\ ,\quad -{\pi\ov 2\l} \ln\left({1+\n\ov 1-\n}\right)
\leq t-t_0 < \infty\ ,
\label{odcc}
\ee
where $t_0$ denotes again an arbitrary reference scale. We see that in the UV
there is a fixed point at $b=0$ (and $a=1$). The lower bound for $t$ above
is needed for $b$ to stay positive, since only then is \eqn{odcc} a solution 
of \eqn{ac11}. 
For the case of $\n>1$, we have to distinguish the solutions between those 
with $b\leq b_-$ and those with $b\geq b_+$. In the former case we obtain
\be
{1\ov b(t)} = \n + \sqrt{\n^2-1} \cosh{\l \ov \pi} (t-t_0)\ ,\quad
{\rm if }\quad \n>1\ ,\quad t\geq t_0\ ,
\label{jqws}
\ee
and 
\be
{1\ov b(t)} = \n - \sqrt{\n^2-1} \cosh{\l \ov \pi} (t-\tilde t_0)\ ,\quad
{\rm if }\quad \n>1\ ,\quad  {\pi\ov 2\l} \ln\left({\n+1\ov \n-1}\right)
\leq t-\tilde t_0 < \infty\ ,
\label{jqws1}
\ee
where, as before, $t_0$ and $\tilde t_0$ are arbitrary reference scales.
For the trajectory given by \eqn{jqws}, $b$ stays positive. It starts 
at $b=b_-$ for $t=t_0$, and ends at $b=0$ for $t\to \infty$.
For the trajectory given by \eqn{jqws1}, $b$ is always negative and starts 
at $b=-\infty$ for $t-\tilde t_0={\pi\ov 2\l} \ln\left({\n+1\ov \n-1}\right)$ 
and ends at $b=0$ for $t\to \infty$. 
Hence, we see that $b=0$ is a UV fixed point. Also as we 
lower the scale $t$ towards the IR, the solution becomes singular
in both cases. In any case, we then run into non-perturbative regimes.
For trajectories in the region $b\geq b_+$, the solution is still given by
\eqn{jqws1}, but with $-{\pi\ov 2\l} \ln\left({\n+1\ov \n-1}\right)
\leq t-\tilde t_0 \leq 0$.
In the lower limit $b\to \infty$ and in the upper limit $b=b_+$. Hence in
that case we have a singular behaviour of the 1-loop RG
equations towards the IR as well as the UV. As we have mentioned, 
in those cases the corresponding 2-dim field theory is not well defined
at the quantum level and can be considered only as an effective field theory
at scales away from the singularities. 

The 2-dim model corresponding to the fixed point at $b=0$ is obtained 
by setting $g_0=-1$ in \eqn{ddsf}
\ba
ds^2 & = &{ 1 \ov \sin^2\th } 
\ \left(d\th^2 + \sin^2\th d\phi^2\right) \ ,
\nonumber \\
B & = & - 2 i \cot\th\ d\th \wedge d\phi \ .
\label{dsbth11}
\ea
The fact that \eqn{dsbth} approaches a free-field conformal field theory at the
fixed point 
is similar to the case of an integrable model (different from \eqn{dsbth}),
representing also a
1-parameter deformation of $S^2$, that was considered 
in \cite{sausage}. It is interesting to investigate whether or not \eqn{dsbth}
represents also an integrable perturbation of $S^2$.

\section{Concluding remarks}

We have constructed a new class of 2-dim field theories with 
target spaces corresponding to deformations of coset spaces $G/H$. 
Our models correspond to special points of the classical 
moduli space of models related by \PL, 
where a local invariance develops. A classification
of all possible models that arise with such a procedure is an interesting open 
problem and can be done by analyzing the general conditions \eqn{rel1},
or equivalently \eqn{kjd}. By construction these models come in dual pairs.
The corresponding generating functionals depend non-polynomially on the
derivatives of the fields with respect to the space-like variable. 
The latter feature is also manifested 
in an underlying infinite-dimensional algebra with a 
central extension of the parafermionic type.
It would also be interesting to uncover the relation of our models to those 
in \cite{KliSevdressed}.

We have also performed a quite general RG flow analysis using specific 
models with 3- and 2-dim target spaces. As in \cite{PLsfe3}, we conclude
that quantum aspects of the lower dimensional models do not 
necessarily follow by taking the same classical limit as that used to relate
the corresponding 2-dim field-theoretical classical actions.
Concretely, the beta-function equations for the lower-dimensional models 
follow from those of the original models by just setting some parameters 
to their prescribed values (see \eqn{abg1} and \eqn{abg2}). 
However, these values do not necessarily correspond to any
fixed points of the solutions of 
these equations. Using our 3-dim example we saw that in a large domain 
in parameter space, and for a wide range of energies in the UV, 
the description is effectively perturbative 
with a UV-fixed point exactly where the local gauge invariance develops.
We believe that this feature will persist for more general models 
related by \PL.
In that respect it would be very interesting to study the RG
flow in general using \eqn{action1} and \eqn{action2} 
and possibly to formulate this flow in a duality-invariant way.

\bigskip\bigskip

\centerline{\bf Acknowledgements}

I would like to thank P. Forgacs for interesting discussions related to 
renormalization group flow in 2-dim field theories 
and for bringing \cite{sausage} into my attention. I also thank
C. Zachos for an e-mail correspondence.

\bigskip\bigskip 

\centerline{\bf Note added}

I would like to thank C. Klimcik and the referee for informing me  
that the models constructed here have been studied before, 
at a purely classical level, in \cite{KliSevdressed}. 
The present paper provides an alternative, more natural to physicists, 
complementary view point on the classical origin
of these models. Moreover, we have further elucidated their structure 
by providing explicit examples and 
exploring quantum aspects of the renormalization group flow.

\appendix 
\section{ Proof of \eqn{kjd} }
\setcounter{equation}{0}
\renewcommand{\theequation}{\thesection.\arabic{equation}}

In this appendix we prove that \eqn{kjd1} (or equivalently \eqn{kjd})
follows from the conditions \eqn{rel2}) (or equivalently \eqn{rel1}). 
First we rewrite \eqn{kjd1}, using an obvious matrix notation, as
\be
S\a\inv - \a^T S = 0\ , \qq  A \a\inv -\a^T A = -b^T\ .
\label{kj1}
\ee
In \cite{PLsfe2}
explicit expressions for the matrices $a,b,\Pi$ were found in terms of
normal coordinates parametrizing the group manifolds.\footnote{We 
also correct a misprint in the expression for the matrix $\tilde b$ as 
it appeared in \cite{PLsfe2}. 
In eq. (41), $n!$ should be replaced by $(n+1)!$.}
In our case 
we have $h=e^{i x^a T_a} \in H \subset G$. Then defining two matrices $f$ and 
$\tilde f$ with matrix elements
\be
f_\a{}^\b = f_{\a c}{}^\b x^c\ ,\qq  \tilde f^{\a\b} =\tilde f^{\a\b}{}_c x^c
\ ,
\label{parr}
\ee
we obtain
\be
\a = (e^{-f})_\a{}^\b\ ,\qq b = \sum_{n=0}^\infty \sum_{m=0}^n
{(-1)^m\ov (n+1)! } (f^T)^{n-m} \tilde f f^m\ .
\label{parr1}
\ee
Using these expressions it is easy to show that proving \eqn{kj1} 
is equivalent to proving 
\ba
&& S f^{n+1} + (-1)^n (f^T)^{n+1} S = 0\ ,\qq n\geq 0\ ,
\nonumber \\
&& A f^{n+1} +(-1)^n (f^T)^{n+1} A =
\sum_{m=0}^n (-1)^m (f^T)^m \tilde f f^{n-m} \ ,\qq n\geq 0 \ .
\label{ghj}
\ea
Their proof proceeds by induction. For $n=0$, the above conditions reduce
to 
\be
S f + f^T S = 0 \ , \qq A f + f^T A = \tilde f\ .
\label{re2}
\ee
These are nothing but the conditions \eqn{rel2} (in a matrix notation
after we contract by $x^c$ appropriately) and by assumption they are 
satisfied.
Asumming that \eqn{ghj} are valid for $n=m$ for some $m\geq 1$, we
may easily show, with the aid of \eqn{re2}, that they also hold for $n=m+1$.
That proves \eqn{ghj} for all $n\geq 0$.


\end{document}